\documentclass[aps,onecolumn,preprint,superscriptaddress,showpacs]{revtex4}

\usepackage{graphicx}
\usepackage{dcolumn}
\usepackage{bm}
\usepackage{color}
\usepackage{amsmath}
\usepackage{amstext}
\usepackage{axodraw}
\usepackage{pstricks}
\usepackage{color}
\newcommand{\lsim}{ \mathop{}_{\textstyle \sim}^{\textstyle <}}

\newcommand{\lsm}{$LSP_{SM}$}
\newcommand{\ldark}{$LSP_{Dark}$}

\newcommand{\be}{\begin{eqnarray}}
\newcommand{\ee}{\end{eqnarray}}

\newcommand{\gev}{\rm \, GeV}

\begin{document}

\title{LHC Signals for a SuperUnified Theory of Dark Matter}
\author{Nima Arkani-Hamed}
\email{arkani@ias.edu}
\affiliation{School of Natural Sciences, Institute for Advanced Study, Einstein Drive, Princeton, NJ 08540}

\author{Neal Weiner}
\email{neal.weiner@nyu.edu}
\affiliation{Center for Cosmology and Particle Physics, Department of Physics, New York University,
New York, NY 10003}

\date{\today}

\begin{abstract}
A new theory of WIMP Dark Matter has been proposed, motivated
directly by striking Data from the PAMELA and ATIC collaborations.
The WIMP is taken to be charged under a hidden gauge symmetry $G_{Dark}$,
broken near the GeV scale; this also provides the necessary
ingredients for the ``exciting" and ``inelastic" Dark Matter
interpretations of the INTEGRAL and DAMA signals. In this short note
we point out the consequences of the most straightforward embedding
of this simple picture within low-energy SUSY, in which $G_{Dark}$
breaking at the GeV scale arises naturally through radiative
corrections, or Planck-suppressed operators. The theory predicts
major additions to SUSY signals at the LHC. A completely generic
prediction is that $G_{Dark}$ particles can be produced in cascade
decays of MSSM superpartners, since these end with pairs of MSSM
LSP's that in turn decay into the true LSP and other particles in
the dark sector. In turn, the lightest GeV-scale dark Higgses and
gauge bosons eventually decay back into light SM states, and
dominantly into leptons. Therefore, a large fraction of all SUSY
events will contain at least two ``lepton jets'': collections of
$n\ge 2$ leptons, with small angular separations and GeV scale
invariant masses. Furthermore, if the Dark Matter sector is directly
charged under the Standard Model, the success of gauge coupling
unification implies the presence of new long-lived colored particles
that can be copiously produced at the LHC.

\end{abstract}


\maketitle

\section{First Hints for Dark Matter and New Physics?}

Recent years have seen a growing body of astrophysical signals
hinting at the existence of dark matter:

\begin{itemize}
\item PAMELA finds an excess of the positron fraction in energies
from $\sim$ 10 $\to$ 50 GeV \cite{boezio}, confirming earlier excesses seen at HEAT \cite{Barwick:1997ig,Coutu:1999ws} and AMS-01 \cite{Aguilar:2007yf}. ATIC sees an excess in $e^+$ or
$e^-$, going all the way out to energies of order $\sim 500-800$ GeV \cite{ATIC2005}.
Finally, the WMAP ``haze" \cite{Finkbeiner:2003im,Dobler:2007wv} can be explained by a similar flux of
$e^+/e^-$ from the galactic center, synchrotron radiating in the
galactic magnetic field, which could arise from dark matter annihilations \cite{Finkbeiner:2004us,Hooper:2007kb}. This interpretation of the WMAP Haze
predicts a large signal for GLAST (now FERMI), from the inverse
Compton scattering of the $e^+/e^-$ off starlight, and will be
tested very soon. At zeroth order, these signals are all consistent
with each other and with an interpretation in the terms of
reasonable Dark Matter candidates annihilating into SM states with a
reasonable annihilation cross-section.

\item The INTEGRAL experiment detects a 511 KeV emission line from
the galactic center \cite{Churazov:2004as,Weidenspointner:2006nu,Weidenspointner:2007}, consistent with the injection of $\sim$ few MeV
positrons. Naively the mass scale here is very different than that
associated with the above anomalies, but if the Dark Sector contains
a number of nearly degenerate states with small splittings $\sim$
few MeV, as in the framework of exciting dark matter, (XDM) \cite{Finkbeiner:2007kk}, then
these positrons could also arise from DM annihilation.

\item The DAMA signal \cite{2008arXiv0804.2741B}, which is still compatible with the null
results of the other DM experiments within the framework of
inelastic dark matter (iDM) \cite{Smith:2001hy,Tucker-Smith:2004jv,2008arXiv0807.2250C}, with $\sim 100$ KeV splittings between
dark matter states, not too much smaller than the splittings already
required by XDM for the INTEGRAL excess.

\end{itemize}

While there may be alternative explanations for some of these anomalies (for instance, pulsar wind nebulae for the local electronic excesses \cite{1995A&A...294L..41A}), the multiple sources, particularly for high energy electrons/positrons both nearby and in the galactic center, invite the consideration of a connection to dark matter. If we do, we are immediately led to a number of qualitative lessons:
\begin{itemize}
\item (0) The most obvious lesson is that there is weak-scale Dark Matter and
it is annihilating into the Standard Model with a sizeable
cross-section. Thus for instance, the DM can't be a gravitino in
low-energy SUSY.

\item (I) All the fluxes resulting from DM annihilation are proportional to
$n_{DM}^2 \sigma_{ann}$. Using typical values for $n_{DM}$, the
PAMELA/ATIC data seem to require a $\sigma_{ann}$ which is $\sim$
100 times bigger than what one would expect from ordinary WIMPS \cite{Cholis:2008hb,Cirelli:2008jk,Cirelli:2008pk}.
Most interpretations so far instead assume that the relevant
$n^2_{DM}$ might be underestimated, but, in our view, such large
``boost factors" seem implausible. Instead, one has to explain how
the annihilation cross-section can be so large. This motivates the
thought the DM is coupled to new light states, with a mass near $\sim
1$ GeV, and that exchange of these states with the slowly moving DM
particles gives a Sommerfeld enhancement needed to boost the
cross-section \cite{TODM}\footnote{The Sommerfeld enhancement was first explored in the context of dark matter in \cite{2005PhRvD..71f3528H,2007PhLB..646...34H}, arising from weak interactions for multi-TeV WIMPs. More recently, it has been connected to signals from ``Minimal Dark Matter,''  \cite{2007NuPhB.787..152C} including PAMELA signals \cite{Cirelli:2008jk}.}. Such light bosons yield annihilation channels that can produce copious leptons without excessive pions and anti-protons \cite{Cholis:2008vb,Cholis:2008qq}. Moreover, this scale is interesting, because a lighter sector might
also play a role in explaining the INTEGRAL and even DAMA signals.

\item (II) The ATIC data in particular suggest that the Dark Matter
particle is at least as heavy as 500-800 GeV \cite{Cirelli:2008pk,paper0}. If, as is common in most
extensions of the Standard Model motivated by naturalness, the Dark
Matter is the lightest state of new physics, having the {\it bottom}
of the spectrum near 800 GeV begins to make the theory very un-natural
indeed. If we want to hold on to the idea of naturalness, it had
better be that the DM is {\it not} the lightest state of new
physics, but instead some state with vector-like quantum numbers
under the Standard Model, which is stable or sufficiently long-lived
as a consequence of a new exact or approximate symmetry.

\end{itemize}

Starting from these qualitative lessons, \cite{TODM} proposed a
simple theory for explaining all the Dark Matter anomalies: the Dark
Matter arises from a multiplet of vector-like states, with some or
all of their flavor symmetry gauged. We show that this
picture for Dark Matter is naturally embedded in extensions of
Standard Model motivated by solving the hierarchy problem and
particularly with low-energy SUSY. Indeed, when this set-up is
embedded in the rubric of low-energy SUSY, it adds two exciting
ingredients to discovery physics at the LHC, associated directly
with supersymmetric cousins of the lessons (I) and (II) above:

\begin{itemize}

\item{SUSY(I)} If there are new light gauge states, it is
reasonable to imagine that the SUSY particles in this new sector are
also much lighter than ours, and thus, the LSP in the other sector
is lighter than ours. Thus, our LSP will decay into the Dark sector.
But, as argued in \cite{TODM}, at least some states in the Dark
sector should decay directly into Standard Model states;  as we will
see, these decays will naturally be predominantly to $e^+ e^-/\mu^+
\mu^-$, and may be associated with displaced vertices. The lepton
pairs will be unusual; their invariant mass will be $\sim$ GeV, and
given that the parent particles in the Dark sector will be very
highly boosted with a $\gamma \sim 10^2$,  in a typical decay the
leptons will be also be produced with tiny opening angles. We'll
refer to such groups of high $p_T$ leptons with small opening angles
and $\sim$ GeV invariant masses as ``lepton jets". Thus, a large
fraction of SUSY events at the LHC should be accompanied with at
least {\it two} ``lepton jets''.

\item{SUSY(II)}
It is possible that the Dark Matter is directly charged under
the Standard Model, or more generally, that there are states
charged under the symmetry that keeps the Dark Matter stable
that are also charged under the Standard Model. If we wish to
preserve the supersymmetric picture of gauge coupling
unification, these states should come in  multiplets that also
contain other colored particles. These colored particles can be
long-lived(though short-lived enough cosmologically).

\end{itemize}

While our discussion is framed within the context of low-energy
SUSY, some of the conclusions hold in a wider class of theories for
new physics. The signals associated with decays into the dark sector
and back follow in any theory with a particle charged under the SM
that is nonetheless stable in the absence of a small coupling to the
Dark Sector, while the new colored states should be expected in any
picture in which the Dark Matter is charged under the Standard Model
and gauge coupling unification is taken seriously.

\section{The Minimal SuperDark Moose}

\begin{figure}
\hskip -0.4 in
\scalebox{0.4}{
\fcolorbox{white}{white}{
  \begin{picture}(100,90)(100,-50)
    \SetWidth{3.0}
    \SetColor{Black}
    \Text(15,50)[lb]{\Huge{\Black{a)}}}
    \CArc(165,63)(84.85,135,495)
    \CArc(165,63)(84.85,135,495)
    \CArc(480,63)(84.85,135,495)
    \Line(422,125)(351,157)
    \Photon(248,60)(395,60){7.5}{8}
    \Text(85,60)[lb]{\Huge{\Black{$SUSY\, breaking$}}}
    \SetWidth{3.5}
    \Line(315,75)(335,45)
    \Line(315,45)(335,75)
    \DashLine(225,125)(293,157){2}
    \Text(47,20)[lb]{\Huge{\Black{$G_{\rm dark}$}}}
    \Text(155,20)[lb]{\Huge{\Black{$MSSM$}}}
  \end{picture}
}
}
\scalebox{0.4}{
\fcolorbox{white}{white}{
  \begin{picture}(100,90)(0,-50)
    \SetWidth{3.0}
    \SetColor{Black}
    \Text(15,50)[lb]{\Huge{\Black{b)}}}
    \CArc(165,63)(84.85,135,495)
    \CArc(165,63)(84.85,135,495)
    \CArc(480,63)(84.85,135,495)
    \Line(250,63)(395,63)
    \Line(422,125)(351,157)
    \Text(85,60)[lb]{\Huge{\Black{$SUSY\, breaking$}}}
    \SetWidth{3.5}
    \Text(47,20)[lb]{\Huge{\Black{$G_{\rm dark}$}}}
    \Text(155,20)[lb]{\Huge{\Black{$MSSM$}}}
  \end{picture}
}
}
\vskip -0.4in
\caption{The minimal supersymmetric model (a) and the minimal SuperDark Moose (b).}
\label{fig:minmoose}
\end{figure}
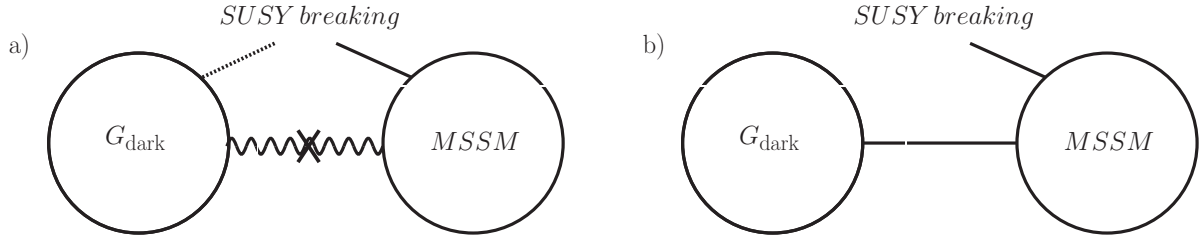

The discussion of \cite{TODM} was mainly concerned with elucidating
a picture of the Dark Matter sector, but ideally this picture should
emerge from a theory that also solves the hierarchy problem. There
are essentially two classes of theories we consider, and they are
shown in figures \ref{fig:minmoose}a,b.

The model with the minimal field content (figure
\ref{fig:minmoose}a), contains no fields in the low energy theory
which are simultaneously charged under both $G_{SM}$ and $G_{Dark}$.
The dark matter must be stable (on cosmological timescales at
least), and this could arise from any range of accidental or exact
discrete symmetries $G_\chi$, global or gauged. We assume the Dark
Matter particle has mass of order the weak scale, while many or all
of the other fields charged under $G_{dark}$ have masses O(GeV), a
scale whose origin we shall come to. A question we must address is
why the dark matter mass is of order the weak scale if the sector
 is largely disconnected from the standard model. This could arise naturally if the Dark Matter mass arises from the same physics
 that sets the MSSM $\mu$ term, linking the scales to each other, for instance via an NMSSM-type mechanism.
We will assume for the moment that the only low-energy connection
between the standard model sector and the GeV-scale particles comes
through a mixing between the dark sector gauge fields and the
standard model gauge fields. Presumably, such a mixing comes from
fields which are charged under both $G_{SM}$ and $G_{Dark}$, but
these may be extremely heavy, even string states.

Given that the kinetic mixing needs some states charged under both
$G_{Dark}$ and $G_{SM}$, a second natural possibility is that there
are ``link" fields in the low-energy theory at the TeV scale, as
in the minimal SuperDark Moose of figure (\ref{fig:minmoose}b). The
links can be neutral under $G_\chi$, in which case they will be
unstable (unless protected by yet another symmetry). If the links
are charged under $G_\chi$, and there are no other $G_\chi$ charged
states charged under $G_{Dark}$, then the lightest of the link
fields will be the Dark Matter. More generally, if there are also
fields charged only under $G_{\chi}$ and $G_{Dark}$, the Dark Matter
will be some linear combination of these states and the link fields,
which can mix after electroweak symmetry breaking.

The idea that dark matter could contain interactions with some new long-distance force has a significant history. The consequences of a new $U(1)$, mixing with hypercharge was first explored in \cite{Holdom:1985ag}, and has been studied extensively within ``mirror dark matter'' \cite{Foot:1995pa}. More recently, forces have been invoked for more phenomenological purposes, in particular in ``exciting dark matter'' \cite{Finkbeiner:2007kk} (which is relevant to our discussion here), ``secluded dark matter'' \cite{Pospelov:2007mp}, MeV-scale dark matter \cite{Boehm:2003hm,Boehm:2003ha}, and WIMPless dark matter \cite{Feng:2008ya}.

The gauge structures in figures \ref{fig:minmoose} in particular, are very similar to those used in \cite{Hooper:2008im,Feng:2008ya}, where the radiative effects were used to generate dark matter at new mass scales, that nonetheless had the relic abundance expected for a WIMP. Here, our dark matter particle is still weak-scale, but the radiative effects will generate mass scales for $G_{dark}$ breaking in a similar fashion. 

As we'll shortly see, the addition of SUSY and SUSY breaking makes
it very natural for the $G_{Dark}$ symmetry to be broken with dark
gauge boson masses at the $\sim M_{Z_{Dark}} \sim \alpha M_Z \sim$
GeV scale. As in \cite{TODM}, this then radiatively induces
splittings between the various DM states of order $\delta M_{DM}
\sim \alpha M_{Z_{Dark}} \sim$ MeV, automatically providing the
necessary ingredients for the XDM and iDM interpretations of the
INTEGRAL and DAMA signals. There are other possible sources of
splittings of the same size. For instance, if the $G_{Dark}$ quantum
numbers of the Dark Matter are such that the first coupling to Dark
Higgses arises from dimension 5 operators (analogously to neutrino
masses in the Standard Model), then if these operators are generated
at the TeV scale, we will get splittings $\sim$ GeV$^2$/TeV $\sim$
MeV as well.

We should emphasize that from a top-down point of view, there is no
particular rationale for these new particles, as they don't in
themselves play an obvious role in solving the outstanding mysteries
of particle theory, such as the hierarchy problem. Having said that,
introducing additional vector-like states charged under another
gauge symmetry is not particularly exotic, and indeed such ``moose"
or ``quiver" structures for gauge theories arise very naturally in
many more complete frameworks for UV physics such as string theory.
At any rate, our motivation for introducing these structures comes
entirely from astrophysical Data and not the desire to engineer
exciting collider phenomenology. Nonetheless, as we will see, this
set-up incorporates all the physics we have discussed while further
providing a natural explanation for why $M_{Z_{Dark}} \sim \alpha
M_Z$ is near the GeV scale. It can also impact LHC collider
phenomenology in a dramatic way.

\subsection{Natural Scale Generation and Low-Energy SUSY Breaking}
We would like to have a natural understanding of why the scale of
$G_{Dark}$ breaking is low. If we have high-scale SUSY breaking
mediated by gravity or its cousins like anomaly mediation, we would
instead  expect that the soft masses in all the sectors are
comparable. Thus we are led to imagine SUSY breaking and mediation
at a lower scale.

Suppose the Dark Matter fields have soft masses $M_S$ of the same
order as the MSSM fields $M_S \sim$ 100's of GeV, and suppose
further that the $G_{Dark}$ sector only get SUSY breaking from DM
loops. Then, we naturally induce SUSY breaking soft masses  at
two-loops
\begin{equation}
M_{SoftDark}^2 \sim (\frac{\alpha}{4 \pi})^2 M_S^2
\end{equation}
in the dark sector, leading naturally to symmetry breaking with
$M_{Z_{Dark}} \sim \frac{\alpha}{4 \pi} M_S$. Therefore, we get the
needed hierarchy of scales with $G_{Dark}$ breaking at the 100 MeV -
1 GeV scale naturally \footnote{A similar approach was used in
\cite{Hooper:2008im} to achieve MeV-scale dark matter.}.

Why would the dark matter have soft masses of $O(100 GeV)$? If we are assuming that whatever generates the $\mu$ term for the Higgsinos is responsible for the scale of the dark matter mass, then it is natural for it to generate a $B_\mu$ term as well, in which case the dark matter fields serve as SUSY breaking messengers for $G_{Dark}$, generating soft masses $O(\alpha_{Dark} m_{SUSY}/ 4 \pi$. If the dark matter mass scale is an accident, or, somehow generates a weak scale mass without a $B \mu$ term, it is possible for supersymmetry breaking to be transmitted to the dark sector through non-renormalizable operators.

This setup is extremely natural if we take the Dark Matter to be
charged under the SM as well, within the context of some low-energy
SUSY breaking scenario, such as gauge mediation. In this case, the
DM would also pick up a ``$\phi^*\phi"$ soft mass of the same order
as the other MSSM fields, but the dark states uncharged under the SM
would only get soft masses from DM loops. Since the gravitino is the
LSP, if we wish to preserve WIMP dark matter we need some extra
fields in any case, and gauge mediation makes a cascade of
radiatively generated scales very natural.

As we have already mentioned, the mass of the DM particle can be
fixed to near the weak scale by the same mechanism that fixes $\mu$
to be near the weak scale. A popular way of doing this is through
the addition of a singlet field $S$ as in the NMSSM, it is then
natural to have $S$ couple in the superpotential $\kappa S H_u H_d +
\kappa^\prime S F F^c$. Then the vev of $S$ will determine both
$\mu$ and the mass of the Dark Matter particle. It is amusing to
note that usually in gauge mediation, it is difficult to make the
NMSSM work in detail, since $S$ fails to get a large enough negative
soft mass from simply coupling to $H_u H_d$. However an additional
coupling to some $(5 + \bar 5)'s$ can increase this negative
mass$^2$ significantly and give a viable solution to the $\mu$
problem \cite{Dine:1993yw}.

Even the dark matter does not acquire a large soft mass, and is not
charged under the SM at all,  we can still get the $\sim$ GeV scale
for the soft masses in a natural way in the context of quite
high-scale gauge mediation, with the $\sim$ GeV gravitino mass, and
a generic $\sim$ GeV size gravity-mediated SUSY breaking. This is
about the largest magnitude tolerable for comfortably solving the
MSSM flavor problem, since the flavor splittings are $\sim \delta
m^2/m^2 \sim$(1 GeV/100 GeV)$^2 \sim 10^{-4}$. It also represents a
reasonably natural messenger scale close to the GUT scale. But while
this ``Planck slop" is a small perturbation in the MSSM sector, it
can generate $\sim$ GeV soft masses in the Dark sector, leading
again to $G_{Dark}$ breaking at the GeV scale.

Finally, if there are link fields with masses and soft masses in the
near the $\sim$ TeV scale, then they will also act as ``messengers"
for the $G_{Dark}$ sector, again generating dark soft masses at
two-loops, naturally near the $\sim$ GeV scale.

\subsection{Shedding light on $G_{Dark}$}

Let us now examine how the $G_{Dark}$ fields communicate with the
MSSM sector. The success of BBN tells us that we shouldn't have any
massless states in the dark sector; it is most natural to assume
that all the lightest new states have a mass in the same $\sim$ GeV
range, and that they can only decay back to SM states. Indeed, this
is necessary for the interpretation of the PAMELA/ATIC data given in
\cite{TODM}, since we assume that the DM annihilations primarily
occur into the light states in the dark sector, and these must decay
back to $e^{+} e^{-}$ a large fraction the the time to explain the
observed signals. Thus we have to examine the leading interactions
between the dark sector and the SM, and determine how the lightest
states in the dark sector can decay into SM states.

Let's begin by considering how the bosonic states in the new
sector--Higgses and gauge bosons--couple to the SM. These are
necessarily produced in DM annihilation, and this part of our
discussion holds generally for {\it any} version of the scenario in
\cite{TODM}, whether or not it is supersymmetric.
The Lagrangian for these theories is of the form
\begin{equation}
{\cal L} = {\cal L}_{SM} + {\cal L}_{Dark} + {\cal L}_{mix},
\end{equation}
and we wish to determine the leading interactions possible between the Dark and SM sectors ${\cal L}_{mix}$.

For simplicity, let's start by imagining that the new gauge sector
has a $U(1)_{Dark}$ symmetry with gauge field $a_{Dark}^\mu$. Then,
the leading interaction with the Standard Model at low energies is
through kinetic mixing with the photon
\begin{equation}
{\cal L}_{mix} = - \frac{1}{2}\epsilon f^{\mu \nu}_{Dark} F^{\mu \nu},
\end{equation}
where of course this coupling would  have to arise from a mixing with hypercharge at energies above the weak scale.
We also assume that there is a Higgs charged only under $U(1)_{Dark}$, so that in Unitary gauge we get a mass term for $a^\mu_{Dark}$
\begin{equation}
m^2 a^2_\mu.
\end{equation}
It is natural to assume that the kinetic mixing is absent in the
UV--for instance if $U(1)_Y$ is embedded in a non-Abelian GUT at
some scale. Then the mixing can be generated radiatively by loops of
particles--that may include the Dark Matter itself--that are charged
under both sectors; this gives $\epsilon\sim 10^{-3}$ as a
reasonable estimate. Actually, in a completely generic model, one
might expect this mixing to be enhanced by a large logarithm $\sim$
log $(M_{GUT}/M)$ from a high scale like the GUT scale. However, if
we imagine that these new states fill out complete $SU(5)$
multiplets with colored and uncolored states,  we instead get a
calculable mixing, with log$(M_{GUT}/M)$ replaced by
log$(M_{colored}/M_{uncolored})$. If there are no light states in the theory charged under both $G_{dark}$ and $G_{SM}$, then very heavy fields (near the GUT scale) would also be expected to generate a mixing, except now, the natural scale is {\em two}-loop, or $\epsilon \sim 10^{-6} - 10^{-4}$, because the split GUT multiplets have only a log-enhanced splitting at low energies.

We can study the physics conveniently by making the  the field redefinition $A^\mu \to A^\mu + \epsilon a^\mu_{Dark}$, which removes the kinetic mixing (and also change the
$a^\mu_{Dark}$ kinetic terms by an irrelevant  $O(\epsilon^2)$ amount). We thus induce a coupling between the electromagnetic current and $a^\mu_{Dark}$:
\begin{equation}
\epsilon a_{Dark}^{\mu} J^{EM}_\mu.
\end{equation}
Note that this does not imply that the dark matter carries electric charge,
as is of course guaranteed by gauge invariance.
The linear combination of gauge fields which is Higgsed is precisely that combination which couples to dark matter,
while the independent, massless combination couples only to standard model fields.

More generally, we can imagine a non-Abelian $G_{Dark}$, where the
dimension 4 kinetic mixing is absent. We can still can get an S-
parameter type operator mixing to the photon if there are particles
that couple to the other sectors Higgs and the SM; this will give us
kinetic mixing operators of the same form but effectively
suppressing $\epsilon$ by a factor of $(v_{Dark}/M)^p$ where $p$
depends on the Higgs quantum numbers. For instance if the Higgses
$\Phi_{Dark}$ are in the adjoint, we can have operators of the form
$\frac{1}{M} Tr (\Phi_{Dark} f^{\mu \nu}_{Dark})F_{\mu \nu}$ with
$p=1$, while e.g. for an SU(2) gauge theory with doublet Higgses,
the analog of the usual S-parameter operator would have $p=2$ etc.

Going to Unitary gauge we will have a collection of Dark Gauge
fields $a^\mu_{Dark\, i}$, and  we'll have
\begin{equation}
m^2_{i j} a^\mu_{Dark\, i} a_{\mu\,Dark\, j} - \frac{1}{2}\epsilon_i f^{\mu \nu}_{Dark\, i} F^{\mu \nu},
\end{equation}
where the $\epsilon_i$ are naturally $\sim 10^{-3}$ for an Abelian factor and are further suppressed for the non-Abelian factors. Redefining $A^\mu \to A^\mu + \epsilon_i a^\mu_{Dark\, i}$, ignoring the tiny $O(\epsilon^2)$ kinetic term corrections for the $a_i$, and changing to mass eigenstate basis
by diagonalizing $m^2_{ij} = U^\dagger_{iI} m_I^2 U_{Ij}$, we get a coupling to the electromagnetic current
\begin{equation}
\epsilon_I a^\mu_{Dark\, I} J^{EM}_\mu, \, \, {\rm with\,} \epsilon_I = U_{I i} \epsilon_i.
\end{equation}
Note that if a $U(1)$ factor is present, in  general there will be
Higgses charged under both $U(1)$ and non-Abelian factors, so
$m^2_{ij}$ will mix all the spin one particles and the mass
eigenstates can all have sizeable $U(1)$ components. Thus, even if
there is a single $U(1)$ factor, we can get $O(10^{-3})$ size
couplings of all the massive gauge bosons to the electromagnetic
current.

If this spin-1 particle can't decay in its own sector for any
kinematical reason, then it will decay through this coupling to the
SM. If it is lighter than $\sim 1$ GeV, then it can't decay to
protons and anti-protons, while it could decay to $K^+ K^-,\pi^+
\pi^-,\mu^+ \mu^-$ and $e^+ e^-$. This is is encouraging because a
huge fraction of these events end up having $e^+ e^-$, and very few
prompt photons from $\pi^0$ decays, so this is the range we prefer
to give the maximal enhancement to the PAMELA/ATIC signal without
polluting other channels. Note that the decay length is
\begin{equation}
\tau \sim (\alpha \epsilon^2 m_{Z_{Dark}} N_{decay channels})^{-1}
\sim (\frac{10^{-7}}{\epsilon})^2 cm
\end{equation}
So, in the case where the mixing between the sectors arises only
after dark sector symmetry breaking, these decay lengths can be
macroscopic, but otherwise the decays are prompt.

Let us consider the Higgses in the dark sector. They necessarily
have an interaction of the form $m_{light} h a_\mu^2$ with heavy
gauge bosons, we have a variety of possibilities for decays. If the
Higgs is heavier than twice the mass of the lightest spin-1
particle, it will decay to them on-shell, which in turn decay to
leptons, giving rise to remarkable $4$ body decays like e.g. $h \to
e^+ e^- \mu^+ \mu^-$. Alternatively, If the lightest state is spin 1
but the Higgs is lighter than twice this mass, then we will get a
decay of $h$ to the lightest spin 1 plus a single current suppressed
only by one power of $\epsilon$. Finally, if the dark Higgs cannot
decay to any on-shell particles, it will decay through loops of the
dark gauge fields and SM leptons to two leptons, with a width
suppressed by $\epsilon^4/({16 \pi^2})^2$, and through off-shell
gauge bosons with a parametrically similar width, but with four
leptons in the final state. Note that these decays do have a
macroscopic length even for $\epsilon \sim 10^{-3}$.

If the Higgs mixes with the standard model through an operator $\kappa \phi^* \phi h^* h$, it could decay directly into a variety of hadrons if it is heavy enough, or directly to muons, such as described in \cite{Cholis:2008vb}. The mixing angle with the standard model Higgs should not be naturally larger than $10^{-6}$ for completely natural parameters \cite{Cholis:2008vb,2008arXiv0805.3531F}, which would make direct decays to SM fermions possible (i.e., bypassing the intermediate dark gauge bosons). However, in SUSY, this would arise from the presence of singlet or non-renormalizeable operator (for instance arising from such a singlet). In this case, we would expect this to be additionally suppressed. However, we note some decays to pions, kaons and other light hadrons is possible.

Finally the dark sector might also include some light
pseudo-goldstone bosons $\pi$. If the SM fermions carried a charge
under the broken symmetry, then $\pi$ will decay to the heaviest
allowed particle, otherwise it's decays are to two photons; the couplings are
\begin{equation}
\frac{m_\Psi}{f} \pi \bar \Psi \Psi, \, \frac{\alpha}{4 \pi} \frac{\pi}{f} F \tilde{F}.
\end{equation}
Especially in the most natural case where the Dark Matter mass
arises from the symmetry breaking associated with the pseudo, we
should expect $f$ near the weak scale, and in both of these cases
the decay lengths can be macroscopic.

Note that making the dark sector supersymmetric adds new kinds of
particles to the dark sector not explicitly considered in
\cite{TODM}: the DM superpartners, as well as gauginos and new
fermions of the $\sim$ GeV scale dark sector. The new light
particles in particular could in principle provide new DM
annihilation channels, though these would involve exchange of the
heavier DM superpartner and would be suppressed, and more
importantly, annihilation to fermions is chirality suppressed, so
these will have a very small branching ratio relative to the dark
gauge boson channels. Also, in order to have at least some of the
vector bosons primarily decay to leptons as required by ATIC/PAMELA,
we must ensure that they don't decay into the new dark fermions;
this could happen if some of the vectors are lighter than twice the
fermion masses, which is perfectly reasonable. Further aspects of
the mixing between superpartners in the MSSM and Dark sectors are
described in next section.

\subsection{Experimental Limits on Light Gauge Bosons}

In our theory we have light $\sim$ GeV gauge bosons with a tiny
coupling to the electromagnetic current; the current experimental
limits on such particles (dubbed ``U bosons") are discussed and
summarized in \cite{Boehm:2003hm,Boehm:2003ha}. Not surprisingly,
because this coupling doesn't break any of the approximate
symmetries of the Standard Model, the constraints are mild. The
strongest constraint comes from the the 1-loop contribution of this
particle to the muon $(g-2)_\mu$, which is of order
\begin{equation}
\delta (g-2)_\mu \sim {\epsilon^2}\frac{\alpha}{\pi}
\frac{m_\mu^2}{m_{Z_{Dark}}^2} \sim 10^{-11}
\end{equation}
even for $\epsilon \sim 10^{-3}$ and $m_{Z_{Dark}} \sim$ 1 GeV. The
sign is the same as the current (small) disagreement between the
measured value of $(g-2)$ and the SM.

The production of this new gauge boson in any processes is
suppressed by a factor of $\sim \epsilon^2$, and there is always a
background from the same process replacing the on-shell gauge boson
by an off-shell photon. Nonetheless one can get an interesting
signal since the new gauge boson has a miniscule width, much smaller
than any experimental resolution. The best limits discussed in
\cite{Borodatchenkova:2005ct} come from low-energy $e^+ e^-$ machines; best of all
(because of the largest integrated luminosity) from B-factories. The
$U$ boson in produced in $e^+ e^- \to \gamma U$, with $U$ decaying
back to $e^+ e^-$ (though the analysis is essentially the same for
$U \to \mu^+ \mu^-$ as well). Binning the data as a function of the
$m^2_{e^+ e^-}$, the signal would be an excess over the Standard
Model background in a single bin; for energies near $\sim$ GeV,
energy resolutions $\sim$ MeV possible. The analysis of \cite{Borodatchenkova:2005ct}
concludes that a limit $\epsilon \sim 10^{-3}$ could be reached from
B-factory data. However, to our knowledge, this search has not been
done by any of the collaborations. Needless to say it would be
extremely interesting to perform this analysis! If there are no
signals in the current data, an increase of $\sim 100$ luminosity at
a super-B factory can push the limit on $\epsilon$ down by another
order of magnitude to $\epsilon \sim 10^{-4}$. It would be
interesting to explore other experimental probes of such light,
weakly coupled gauge bosons more systematically.

Note that, as pointed out in \cite{TODM}, a coupling $\epsilon \sim
10^{-3} - 10^{-4}$ accompanied by $\sim 100$ KeV splittings amongst
Dark Matter states, can explain the DAMA signal. It is intriguing
that in this same range we get an interesting contribution to
$(g-2)_\mu$, as well as the possibility to detect direct production
of the new gauge bosons.

\subsection{The Early Universe and the GeV-scale spectrum}
Our focus here is on the LHC phenomenology, so we shall not attempt to describe to complete features of the early universe phenomenology. Rather, we are interested in understanding what the implications of freezeout are on the possible decay chains through the GeV-scale $G_{Dark}$ sector.

\begin{figure}
\scalebox{0.5}{
\fcolorbox{white}{white}{
  \begin{picture}(120,50)(40,-20)
    \SetWidth{1.5}
    \SetColor{Black}
    \DashLine(105,91)(180,16){10}
    \DashLine(180,16)(105,-59){10}
    \Photon(180,16)(270,16){7.5}{5}
    \Line(255,31)(285,1)
    \Line(255,1)(285,31)
    \Photon(270,16)(360,16){7.5}{5}
    \ArrowLine(435,-59)(360,16)
    \ArrowLine(360,16)(435,91)
    \Text(10,35)[lb]{\Huge{\Black{$h_{Dark}$}}}
    \Text(10,-25)[lb]{\Huge{\Black{$h_{Dark}$}}}
    \Text(160,35)[lb]{\Huge{\Black{$e^-$}}}
    \Text(160,-25)[lb]{\Huge{\Black{$e^+$}}}
  \end{picture}
}
}
\caption{Process contributing to thermal equilibrium of the $G_{Dark}$ sector.}
\label{fig:eq}
\end{figure}
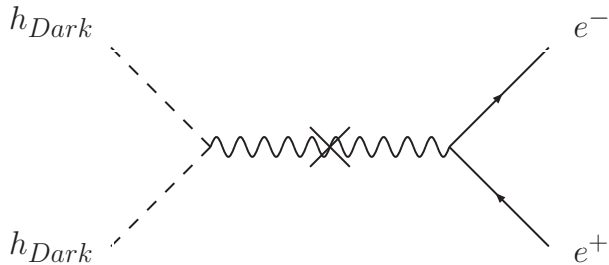

The dark matter will stay in equilibrium, either via annihilations to $G_{Dark}$ gauge bosons (in analogue to the XDM scenario freezeout \cite{Finkbeiner:2007kk})\footnote{Such freezeout has been termed ``secluded" dark matter models \cite{Pospelov:2007mp}, and similar phenomena occur in ``WIMPless'' models \cite{Feng:2008ya}.}, or via annihilations to $G_{Dark}$ and $G_{SM}$ gauge bosons in the case that the DM carries a SM charge as well. Thus, we focus on the equilibrium properties of the light particles. Thus we consider the thermal properties at GeV temperatures, long after the much heavier dark matter has frozen out. The dark Higgses and dark gauge bosons will generally stay in thermal equilibrium with the standard model via s-channel dark gauge boson exchange (figure \ref{fig:eq}). This process will proceed with a cross section $\sigma \sim \alpha^2 \epsilon^2/{\rm GeV}^2$. This will maintain equilibrium between the sectors until the dark particles become non-relativistic.

If the LSP of the dark sector, which we refer to as LSP$_{Dark}$,  is the true LSP (i.e., lighter than the gravitino, as might occur in high-scale gauge mediation), we must check whether it is overproduced in the early universe, but the abundance is easily small enough. For instance, if the $LSP_{Dark}$ is a dark gaugino, t-channel dark Higgsino exchange will allow annihilations into dark Higgses, with $\alpha^2/{\rm GeV}^2$ cross sections, giving a present abundance $\sim 10^{-4}$ times critical density.

\section{Cascade Decays into the SuperDark ``Hidden Valley"}
\label{sec:cascade}
If there are many particles which are kinematically accessible to the LHC, we must ask: how could such $G_{Dark}$-charged states be produced? There are two simple possibilities: we can produce the $G_{Dark}$-charged states directly, or we can cascade into them. We will begin with the latter case.

The presence of a new sector of light particles weakly coupled to
the Standard Model can have dramatic implications for collider
physics, as has especially been explored in recent years by
Strassler and collaborators
\cite{Strassler:2006im,Strassler:2006ri,Strassler:2006qa,Han:2007ae,Strassler:2008fv}.
This is particularly the case in low-energy SUSY with unbroken
R-parity, since in this case the LSP can reside in the new sector,
so that all SUSY events eventually ending up with MSSM LSP's decay
to the new sector. This was discussed at length in the context of
supersymmetric ``Hidden Valley'' theories in
\cite{Strassler:2006qa}.   Here we outline what this physics looks
like in our case; the principle difference between this scenario and
previous studies of ``Hidden Valley'' phenomenology is that the
particular leading interaction between our sector and the dark
sector arises through kinetic mixing with the photon. This has
important implications for the collider phenomenology, for instance,
we do not expect the hidden sector to dominantly decay to heavy
flavor Standard Model states.  Moreover, because we are motivated by
the electronic excesses at PAMELA and ATIC, the mass scale we single
out kinematically favors leptons in final states.

We preface this discussion with a comment: since the Dark Matter is
not the LSP, R-parity is not needed to keep the Dark Matter stable,
and instead another discrete symmetry must be invoked. However,
R-parity is still the simplest explanation for the absence of $B$
and $L$ violating couplings in the MSSM, so we will continue to
assume it is a good symmetry as the simplest possibility, though we
will have a few words about R-parity violation as well.

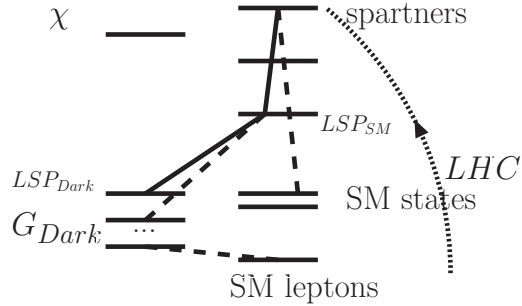
\begin{figure}
\scalebox{0.5}{
\fcolorbox{white}{white}{
  \begin{picture}(100,100)
    \SetWidth{3.5}
    \SetColor{Black}
    \Line(0,180)(60,180)
    \Line(0,40)(60,40)
    \Line(0,60)(60,60)
    \Text(7,11)[lb]{\Huge{\Black{...}}}
    \Line(0,20)(60,20)
    \Line(100,200)(160,200)
    \Line(100,10)(160,10)
    \Line(100,120)(160,120)
    \Line(100,160)(160,160)
    \Line(100,50)(160,50)
    \Line(100,60)(160,60)
    \DashLine(130,200)(145,60){10}
    \Line(130,200)(120,120)
    \DashLine(120,120)(30,40){10}
    \DashLine(30,20)(130,10){10}
    \Line(120,120)(30,60)
    \Text(-15,65)[lb]{\Huge{\Black{$\chi$}}}
    \Text(-25,22)[lb]{\Large{\Black{\ldark}}}
    \Text(-25,8)[lb]{\huge{\Black{$G_{Dark}$}}}
    \Text(64,65)[lb]{\Huge{\Black{spartners}}}
    \Text(64,17)[lb]{\Huge{\Black{SM states}}}
    \Text(33,-8)[lb]{\Huge{\Black{SM leptons}}}
    \Text(57,37)[lb]{\Large{\Black{\lsm}}}
    \Text(90,25)[lb]{\Huge{\Black{$LHC$}}}
    \DashArrowArc(0,0)(260.02,0,50){2}

  \end{picture}
}
}
\caption{Cascade decays into the $G_{Dark}$ sector and lepton jets. The final decay to leptons can arise at the end of a variety of chains in the $G_{Dark}$ charged sector.}
\label{fig:cascade}
\end{figure}

We begin with a bit of nomenclature. We refer to the lightest R-odd particle of the MSSM as $LSP_{sm}$. Similarly, we refer to the lightest R-odd particle charged under $G_{Dark}$ as \ldark . A priori, we make no assumption as to whether the gravitino is the ``true'' LSP or not. As it will happen, the phenomenology is most generically interesting when the gravitino is the true LSP, as in low-scale gauge mediation, although much interesting phenomenology can arise even if the \ldark\, is the LSP.

The basic picture of the phenomenology is shown in figure \ref{fig:cascade}. We assume that LHC SUSY production occurs as in a standard SUSY scenario. This proceeds to cascade down to the \lsm\, and visible matter. If the SUSY breaking scale is sufficiently high (which we shall argue should generically be the case), then the \lsm\, must decay through the hidden sector to reach the \ldark. Because the connection to the standard model goes through the gauge mixing, the states heavy enough to decay to the dark gauge bosons will do so, and those that are lighter will proceed through loops or off shell dark gauge bosons to produce leptons as well. As a consequence, a generic feature of the decay will be ``lepton jets,'' i.e., sets of $n\ge 2$ highly boosted leptons with low ($\sim$GeV) invariant mass.

\subsection{Decay Zoology}

Although the myriad possibilities cannot be listed exhaustively, we attempt to discuss the most important features. The same physics that gives
rise to the $\epsilon$ kinetic mixing between the hidden gauge field
and the photon, will give rise to a mixing between the hidden
gaugino $\eta$ and $\chi_0$
\begin{equation}
\epsilon^\prime \bar \eta \bar \sigma^\mu \partial_\mu \chi_0,
\end{equation}
and so the $\chi_0$ will have a small mixing with the dark sector. It is important to note that in most MSSM models, the gauginos are relatively pure states (i.e., not significant mixtures with the Higgsinos). This is because the mixing terms arising from Higgs vevs are generally much smaller than MSSM SUSY breaking masses (a reflection of the well-appreciated tuning necessary for the MSSM Higgs sector). This is not expected to be the case in $G_{Dark}$, where most likely the fermionic states will be large mixtures of dark-Higgsino and dark-gaugino.

Let us begin with the case where the \lsm\, is a gaugino. In this case, because of the mixing term, we expect a decay such as $\chi_0 \rightarrow h_{Dark} \chi_{Dark}$, where $\chi_{Dark}$ may or may not be \ldark. Subsequently, we will have $h_{Dark}$ decay to leptons, either through on-shell, off-shell, or loops of, dark gauge bosons. If $\chi_{Dark}$ is not the \ldark, we expect it to decay to \ldark\, via on- or off-shell dark gauge boson emission (such as in the case of non-Abelian $G_{Dark}$),  If the gravitino is the true LSP, we expect the \ldark\, to decay further to $\tilde \psi_{3/2}$ and a dark gauge boson or dark Higgs, which then subsequently will decay to additional leptons.

Alternatively, we can consider a case where the \lsm\, is a sfermion $\tilde f$. In this case, we will have $\tilde f \rightarrow f \eta_{Dark}$, where $\eta_{Dark}$ is one of the mixed gaugino-Higgsino states of $G_{Dark}$. If $\eta_{Dark}$ is the true LSP, then this will appear similar to gauge mediation. However, we still expect some decays to states $\eta_{Dark}$ which are not the \ldark\, (of the dark Higgsino/gaugino mixture), which then decay to the \ldark . If we are in a scenario such as low-scale gauge mediation, then, again, we have \ldark $\rightarrow a_{Dark} \tilde \psi_{3/2}$, followed by $a_{Dark} \rightarrow leptons$, or, possibly, further cascades in the situation of \ldark $\rightarrow h_{Dark} \tilde \psi_{3/2}$, or some other state which decays further in the $G_{Dark}$ sector.

In this discussion we have ignored the possibility that the \lsm\,
could instead decay straight to the gravitino. However, this decay
width is of order $m_{\chi_0}^5/F^2$; even for the lowest imaginable
scale $F \sim (10 TeV)^2$, this is subdominant for $\epsilon >
10^{-6}$. If matter/R parity is broken, then there is also a competing $R_p$
violating decay of the \lsm\, to SM particles. As usual with $R_p$
violation, we have to imagine that we are either preserving baryon
number or lepton number. If we are allowing the $qld,lle$ operators,
then their size is constrained minimally by not generating too-large
neutrino masses at 1-loop; this makes the couplings small enough
that for $\epsilon \sim 10^{-3}$, the \lsm\, would still prefer to
decay into the new light sector. Only the purely $udd$ operators
involving all third generation fields can have reasonable
coefficients and significantly depress the decay of \lsm\, into the
new sector.

Clearly there are many more combinatorial possibilities one could envision; we have engaged in this brief discussion here only to make it clear that regardless of the identity of
LSP$_{SM}$, SUSY events will lead to decays into the Dark sector, and these will in return decay back into leptons in our sector, which we now turn to.

\subsection{``Lepton Jets"}

Usually every SUSY event ends with two LSP's plus visible particles;
in our case, the \lsm's further decay into the \ldark\, that still
carries away missing energy, but also goes to the lightest R-even
particles that decay back to $e^+ e^-$,$\mu^+ \mu^-$,$\pi^+ \pi^-$
with a large branching fraction. The parent particles in the dark
sector are boosted with $\gamma \sim M_{LSP_{SM}}/m_{Dark} \sim
100$.  The decay lengths are as quoted in the section II, multiplied
by this $\gamma$ factor.  If $\epsilon$ is as large as can be
consistent with the muon $(g-2)$ constraints, the decay will not
leave a sufficiently large displaced vertex, but with any
suppression of $\epsilon^\prime$ displaced vertices are a distinct
possibility. Regardless of the displaced vertices, the lepton pairs
will have a small invariant mass $\sim$ GeV, and in typical decays,
will come out with small angular separation $\sim 1/\gamma \sim
10^{-2}$. Thus essentially all SUSY events should include {\it at least two}
pairs (4 leptons total) of high-$p_T$ opposite-charge light
particles. In cases where leptons are produced more copiously, it
will be difficult to extract resonances from the combinatorial
background. Nonetheless, because the splittings in the sector are
expected to be $\sim$ GeV, we do not expect any reason for the
leptons to be particularly soft, except as arises in multibody phase
space. Thus, we have the possibility of ``lepton jets'': boosted
groups of $n\ge 2$ leptons with low $\lsim \gev$ invariant masses.
Such objects may have some hadronic states in them, for instance if
the vector can decay to charged pions or if dark Higgses arise with
hadronic decay modes as well. Still the hard lepton content will be
much richer than usual jets, and should make them distinctive even
in this case.

Finally note that ordinarily with low-energy SUSY, unless the
electroweak charged states are quite light $\lsim 300$ GeV, it is
not possible to probe their direct production, and instead one has
to rely on cascade decays to them via the colored states. However,
the presence of lepton jets in essentially all SUSY events dramatically reduces
backgrounds, and should allow a probe of direct electroweak
production to higher masses.

\section{Unification, Colored Particles and DM Production}
So far we have considered the phenomenology that arises without link fields, i.e., the moose of figure \ref{fig:minmoose}a. We can now consider the situation with link fields as well.
If we wish to assume that unification is preserved, then we should take the link fields to arise in complete GUT multiplets, including new colored states. These will allow us a new avenue of $G_{Dark}$ production.

Of course the easiest way to have the unification unaltered
``automatically" is to add states in complete multiplets of $SU(5)$
or $SU(3)^3/Z_3$; lets imagine the simplest example, where there are
$N_F$ $5 + \bar 5$'s of $SU(5)$.  These could have a mass in the
neighborhood of the weak-scale by the same mechanism that makes
$\mu$ close to the weak scale. we call the states $F^c = (D^c,L)$
and $F = (D,L^c)$ with the obvious notation, and reserve lower-case
$q,u^c,d^c,l,e^c$ for the SM states.

The link fields could be unstable, decaying through an operator such as $h_{Dark} D d^c$ where $d^c$
is the usual MSSM superfield (of arbitrary flavor) and $h_{Dark}$ is a Higgs field of $G_{Dark}$. This will result in a decay into hard jets and ``lepton jets'', but no missing $E_T$. If the representations of fields under $G_{Dark}$ are such that no renormalizeable operator can be written, it is possible that the lifetime of this could be sufficiently long that it would not decay in the event. This will be similar to the case where we identify the link fields with the dark matter, and we defer that discussion for the moment.

If we produce the superpartner of the link field, $\tilde D$, we can
have decays $\tilde D \rightarrow D \tilde g$ or  $\tilde D
\rightarrow D \tilde g^*$ . $D$ will then yield hard jets and the
$\tilde g$ or $\tilde g^*$ will produce a SUSY cascade as described
in section \ref{sec:cascade}. Alternatively, we could have decays
$\tilde D \rightarrow D \eta_{Dark}$, where $\eta_{Dark}$ is a dark
Higgsino/gaugino mixture, as before. This will produce the ``lepton
jet" from the $\eta_{Dark}$ decay, while $D$ will give jets plus
leptons. Finally, we may have decays $\tilde D \rightarrow \tilde
h_{Dark} d^c$ and $\tilde D \rightarrow  h_{Dark}\tilde d^c$, which
may arise with comparable rates depending on the mass of the $\tilde
d^c$. The former will yield ``lepton jets" and missing $E_T$ through
the $\tilde h_{Dark}$ cascade, while the latter will yield ``lepton
jets" through the $h_{Dark}$ decay, and ``lepton jets" together with
missing $E_T$ through the $\tilde{d}^c$ via the SUSY cascade
described in section \ref{sec:cascade}.

It is important to note that in these cases, we achieve signatures similar to gauge mediation, augmented with ``lepton jets''.
 However, the dark matter particle $\chi$ is not expected to be produced in these cascades unless it has tree-level superpotential
 couplings to the link fields, or some singlet field that arises in some other decay.

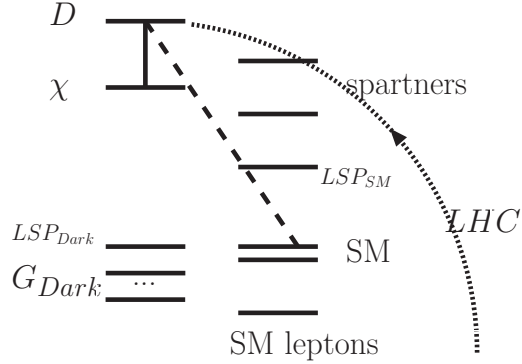
\begin{figure}
\scalebox{0.5}{
\fcolorbox{white}{white}{
  \begin{picture}(100,100)
    \SetWidth{3.5}
    \SetColor{Black}
    \Line(0,230)(60,230)
    \Line(0,180)(60,180)
    \Line(0,40)(60,40)
    \Line(0,60)(60,60)
    \Text(7,11)[lb]{\Huge{\Black{...}}}
    \Line(0,20)(60,20)
    \Line(100,200)(160,200)
    \Line(100,10)(160,10)
    \Line(100,120)(160,120)
    \Line(100,160)(160,160)
    \Line(100,50)(160,50)
    \Line(100,60)(160,60)
    \DashLine(30,230)(145,60){10}
    \Line(30,230)(30,180)
    \Text(-15,80)[lb]{\Huge{\Black{$D$}}}
    \Text(-15,60)[lb]{\Huge{\Black{$\chi$}}}
    \Text(-25,22)[lb]{\Large{\Black{\ldark}}}
    \Text(-25,8)[lb]{\huge{\Black{$G_{Dark}$}}}
    \Text(64,60)[lb]{\Huge{\Black{spartners}}}
    \Text(64,17)[lb]{\Huge{\Black{SM}}}
    \Text(33,-9)[lb]{\Huge{\Black{SM leptons}}}
    \Text(57,37)[lb]{\Large{\Black{\lsm}}}
    \Text(90,25)[lb]{\Huge{\Black{$LHC$}}}
    \DashArrowArc(30,-20)(250.02,0,82){2}

  \end{picture}
}
}
\caption{The simplest diagram of direct production of $G_{Dark}$ sector. Similar diagrams exist for production of $\tilde D$ and for cases where the direct production is of a link field. Such cascades generally produce additional ``lepton jets'', with and without missing energy.}
\label{fig:direct}
\end{figure}

We now consider the alternative possibility, where the link fields
are also charged under $G_{\chi}$. We assume that on these fields,
$G_\chi$ acts as $Z^F_2$, under which $(F,F^c) \to -(F,F^c)$, which
keeps the lightest of these link fields stable. We will continue
assume that the theory has $Z_2^M$ matter parity for the usual
reason of forbidding baryon and lepton number violation, even though
this is no longer necessary to guarantee a stable DM particle, only
briefly addressing the possibility on $Z_2^M$ violation.

Even though the underlying theory only has a $Z_2^F \times Z_2^M$
symmetry, at renormalizable level the low-energy theory has a much
larger $U(1)_L \times U(1)_D$ global symmetry acting on the link
fields, which has important consequences for the phenomenology.

To begin with, note that the DM states here are identical to a
``Higgsino"; as such, there are two degenerate Majorana states, as
guaranteed by the accidental $U(1)_L$ symmetry. We have to assume
that this $U(1)_L$ is broken by higher-dimension operators too, to
give the at least $\sim 100$ KeV splitting needed to avoid having
seen this DM in direct detection experiments arising from couplings
to the $Z$-boson.  For instance we need an operator of the form
\begin{equation}
\label{split1}
 \frac{LL H_u H_u}{M_*}
\end{equation}
which gives a splitting of order $\sim v^2/M_*$; such an operator
can be generated by mixing with a singlet field with mass $\sim M_*$
that could be anywhere from the TeV scale to $\sim 10^{9}$ GeV.

The dark matter will easily annihilate into $W$ bosons, depleting
its relic abundance below the observed level. This provides yet
another motivation for having the dark matter be part of a multiplet
of states; in our picture we of course further gauge some subgroup
of the global flavor symmetry rotating these states into each other.
Note that for the PAMELA/ATIC signals, we need to ensure that the
branching ratio for annihilating into Standard Model $W$'s and $Z$'s
is less than $\sim 10$ \%. It is easy to see that the annihilations
will go into pairs of Dark gauge bosons or Standard Model gauge
bosons; the ratio then scales as the ratios of $({\alpha_{Dark}
C_{Dark}}/(\alpha_{SM} C_{SM})^2$ where $C$ are the Casimirs of the
corresponding representations; even for comparable gauge couplings
this ratio can easily be $\sim 10$.

Note that if the operator of eqn.(\ref{split1}) is generated by
mixing with a Standard Model singlet via the Yukawa coupling $\kappa
LHS$, then $S$ must also be charged under the $G_{Dark}$, and if as
is quite natural it is at the TeV scale, then the Dark Matter will
be some admixture of $L$ and $S$. If the invariant mass of $S$ is
smaller than that of $L$, the DM will be mostly a SM singlet, with
an admixture $\sim \kappa v/M$ of $L$ after electroweak symmetry
breaking. This mixing is naturally $\sim 10$ \%, and very
efficiently suppresses annihilations into SM gauge bosons down to
$\sim 10^{-4}$. However we can get an interesting rate for
annihilations into Higgses, which is controlled by the size of the
Yukawa coupling $\kappa$.

What about the colored partners of the DM? The accidental $U(1)_D$
symmetry guarantees that they are stable at renormalizeable level.
The leading higher-dimension operator that is $Z^F_2$ invariant but
breaks the two separate $Z_2$'s and allows the colored particles to
decay is a dimension five operator
\begin{equation}
\int d^2 \theta \frac{1}{M_G}(D L^c) (d_3^c e_3^c)
\end{equation}
giving a decay width
\begin{equation}
\Gamma \sim \frac{M^3}{32 \pi M_G^2} \sim (.1 s)^{-1}
\end{equation}
making for macroscopically long lifetimes that are however short
enough not to cause trouble with nucleosynthesis. The collider
signals of long-lived colored particles have been studied at length
recently \cite{Arvanitaki:2005nq}; in particular the fact that a
reasonable fraction of them are stuck in the detector, and decay
inside it. One would naturally expect this to arise in our models.
But there is a potentially dramatic addition to this signal. When
the colored particle decays in the detector into its electroweak
partner, it will can also radiate off $G_{Dark}$ bosons that can
promptly re-decay back into SM states. If these include $\mu^+
\mu^-$ pairs, in addition to the usual ``explosion" in the detector
coming from the jets in the colored particle decays, there will be
muons traveling either out towards the muon chamber and/or back
into the tracker!

If we  don't impose matter parity, then the new colored
particles can directly couple to the SM states via for instance a superpotential term,
$\kappa q_3 L D^c$, which would lead to a rapid decay of the colored state to the DM
particle.

Whether or not the colored particle is long-lived, its presence is a
boon for probing the Dark Matter sector at the LHC.Even if the
colored and uncolored states start with a unified mass near the GUT
scale, the colored states will become heavier by a factor of $\sim
2$ in running down to the weak scale. But putting in the ATIC mass
$\sim 800$ GeV, this means that the colored states will have a mass
near $\sim 1.5$ TeV, perfectly accessible to be copiously produced
the LHC. Furthermore, whether with a long or short lifetime, these
decay dominantly directly to the DM particle and not through a
complicated cascade decay process.

\section{Discussion and Outlook}

As we stand on the threshold of the LHC era, astrophysical data
could be giving us a first hint for what is to come. If the
interpretation of \cite{TODM} for PAMELA/ATIC/WMAP
Haze/EGRET/INTEGRAL/DAMA is correct, a host of signals are to be
expected in further astrophysical measurements, beginning with
GLAST/FERMI. As we have argued here, there are also a number of
possible signals at the LHC.

The new dark sector is an example of a ``Hidden Valley", but one whose
properties are motivated and strongly constrained by astrophysical
Data. The unique feature of the ``Hidden Valleys'' that are
motivated by these astrophysical clues is that they have cascades
that end with many leptons, rather than hadronic states. These
``lepton jets'' are the key LHC feature of most any supersymmetric
realization of the theory of dark matter proposed in \cite{TODM}, in
which the dark matter states transform under a gauge symmetry
$G_{Dark}$ broken at the GeV scale.

At the same time, in very natural extensions of these theories,
where the dark matter or other fields in the theory transform under {\em both} $G_{SM}$ and $G_{Dark}$, the requirement of unification promises the possibility of new colored states, some of which can be long lived. Prompt decays of new colored states or superpartners of long-lived colored states can yield topologies with and without missing energy, and with or without ``lepton jets''.

These theories have a rich collider phenomenology, whose complete
analysis would take us well beyond the simple discussion we have
given here.  It is also worth noting that the dark matter
considerations have led us to consider light states in the
neighborhood of $\sim$ GeV, such particles have been considered
purely from a particle physics perspective in the past number of
years, for instance, for the purposes of ``hiding'' the Higgs from
LEP searches \cite{Chang:2008cw}. An intriguing connection can also
be imagined with the 3 strange events observed in the Hyper-CP
experiment \cite{Park:2005eka}, that could be interpreted as a
resonance with a mass just above $2 m_\mu$ decaying to a pair of
muons. And we repeat that it would be interesting to search for
these light gauge bosons in existing B-factory data, as well as in
possible Super-B factories.

 In this short note, we have only sketched the simplest
embedding of the dark matter framework proposed in \cite{TODM} into a more complete supersymmetric picture
of physics beyond the standard model. We leave the important task of
constructing a specific model to future work. However, the sketch
suffices to show that the essential ideas of \cite{TODM} are (A)
reinforced by a SUSY embedding, which can naturally explain the
origin of the lighter GeV dark symmetry breaking scale, and (B) give
rise to dramatic new signals for the LHC, and the possibility of a
direct experimental probe into the rich dynamics of the Dark Sector.
It is now fortunately only a short time before these ideas are
decisively tested experimentally on all fronts.

\vskip 0.2in
\noindent {\bf Acknowledgements}

The authors are very thankful to E.~Witten, P.~Meade, M.~Papucci,
J.~Maldacena, N.~Seiberg, P. Schuster,  L.~Senatore, D.~Shih,
M.~Strassler, N. Toro, T.~Volansky, S.~Thomas, J.~Ruderman,
L.T.~Wang, I.~Yavin, C.~Chung, M.~Baumgart, S.~Rajendran and
P.~Graham for helpful discussions, and especially  I.~Cholis,
L.~Goodenough, D.~Finkbeiner, T.~Slatyer and T.~Vachaspati for
conversations and insightful observations. NW is supported by NSF
CAREER grant PHY-0449818 and DOE OJI grant \# DE-FG02-06ER41417. The
work of N.A.-H. is supported by the DOE under grant
DE-FG02-91ER40654.

\bibliography{superunified}
\bibliographystyle{apsrev}

\end{document}